
\magnification =1200
\baselineskip=13pt

\tolerance=100000

\overfullrule=0pt

\rightline{UR-1406\ \ \ \ \ \ }
\rightline{ER40685-853}

\bigskip

\centerline{\bf SUPERSYMMETRIC THEORIES ON A NON SIMPLY}
\centerline{\bf CONNECTED SPACE-TIME}

\bigskip
\bigskip
\bigskip
\centerline{Ashok Das}
\centerline{and}
\centerline{Marcelo Hott$^\dagger$}
\centerline{Department of Physics and Astronomy}
\centerline{University of Rochester}
\centerline{Rochester, NY 14627}

\vskip 1.5 truein

\baselineskip=18pt

\centerline{\bf \underbar{Abstract}}

\medskip

 We study the Wess-Zumino theory on ${\bf R}^3 \times S^1$ where a spatial
coordinate is compactified.  We show that when the bosonic and fermionic
fields satisfy the same boundary condition, the theory does not develop a
vacuum energy or tadpoles.  We work out the two point functions at one loop
and show that their forms are consistent with the nonrenormalization
theorem.  However, the two point functions are nonanalytic and we discuss
the structure of this nonanalyticity.

\vskip 1.5truein

\noindent $^\dagger$On leave
 of absence from UNESP - Campus de
Guaratinguet\'a, P.O. Box 205, CEP : 12.500, Guaratinguet\'a, S.P., Brazil

\vfill\eject

It is well known that quantum field theories at finite temperature lead to
some novel features.  For example, two point functions as well as higher
point functions become nonanalytic at the origin in momentum space at
finite temperature [1-3].  There are  some cases where such nonanalytic
behavior
does not arise -- the simplest example being that of a self-energy which
involves unequal masses in the loop [4].  One  could possibly also hope for an
improved behavior in a supersymmetric theory where bosonic and fermionic
loops have a tendency to cancel.  Unfortunately, however, supersymmetry is
broken at finite temperature [5-6] and, therefore, this cannot be directly
tested.

On the other hand, studies of quantum fields on non simply connected
space-times have been of interest for quite some time now [7-12].
 The simplest of
such space-times is ${\bf R}^n \times S^1$ which is a prototype of
Kaluza-Klein or string compactification.  Here a space coordinate is
compactified (as opposed to the compactification of the time coordinate at
finite temperature in the imaginary time formalism).  There have been many
studies on such spaces and the calculations, not surprisingly, are
remarkably similar to the finite temperature ones.  The difference,
however, is in the fact that the compactification length has no relation to
the physical temperature and, therefore, one is not required to impose
periodic boundary conditions on bosonic fields and antiperiodic ones on the
fermion fields.
  (Namely, the generating functional is not a
partition function and as a result, the KMS [13] condition need not be
satisfied.)
  Consequently one can require both the bosonic and the fermionic fields to
satisfy either periodic or antiperiodic boundary conditions in the hope
that supersymmetry is maintained.  In this letter, we study a
supersymmetric theory -- the Wess-Zumino theory -- on ${\bf R}^3 \times S^1
$ where one space coordinate is compactified.  We show that only when  both
the bosonic and the fermionic fields satisfy the same boundary condition,
the theory does not develop a zero point energy, consistent with
supersymmetry, which is quite distinct from the behavior of the
nonsupersymmetric theories [9] on such a space-time.  We study the two point
functions for the bosonic and the fermionic fields at one loop and show
that, for either of the boundary conditions (periodic or antiperiodic),
they have the required form consistent with the nonrenormalization
theorem [14].
 However, the two point functions are nonanalytic.  We discuss the
structure of the nonanalyticity and present a short conclusion.

The Wess-Zumino theory [14] is described by the Lagrangian density
$\big($our metric
is
$(+,-,-,-) \big)$
$$\eqalign{{\cal L} = &{1 \over 2} \ \partial_\mu A \partial^\mu A +
{1 \over 2}\ \partial_\mu B \partial^\mu B - {m^2 \over 2}\ \big( A^2 + B^2
\big) + {i \over 2}\ \overline \psi \rlap \slash{\partial} \psi
-{m \over 2}\ \overline \psi \psi\cr
&- g \overline \psi \big( A - i \gamma_5 B \big) \psi - mgA
\big( A^2 + B^2 \big) - {g^2 \over  2} \ \big(
A^2 + B^2 \big)^2\cr}\eqno(1)$$
where $A$ and $B$ are scalar and pseudoscalar fields while $\psi$ is a
Majorana spinor.  Just to recapitulate some of the salient features of this
theory, we note that this theory (in the four dimensional Minkowski
space-time) has zero vacuum energy.  The tadpoles in the theory vanish and
the two point functions have the form
$$\eqalign{\qquad\qquad\qquad &= a \big( k^2 + m^2 \big)\cr
\noalign{\vskip 6pt}%
\qquad\qquad\qquad &= a\big( k^2 + m^2 \big)\cr
\noalign{\vskip 6pt}%
\qquad \qquad\qquad &= a \rlap\slash{k}\cr}\eqno(2)$$
This shows that the theory needs only a single wave function
renormalization constant and this is the essence of the nonrenormalization
theorem.  (Similar results also follow for the vertex functions, but we
will not be concerned with them.)

Next, let us consider the Wess-Zumino theory on ${\bf R}^3 \times S^1$.  We
assume that the $x^3$ coordinate is compactified with a compactification
length $L$.  Thus, the momentum along this axis will be discrete and
depending on the boundary condition will have values
$$\eqalign{p^3_{\rm Per.} &= {2 \pi n \over L}\cr
&\qquad\qquad\qquad\qquad\qquad\qquad n = 0, \pm 1, \pm 2 , \pm 3, \dots\cr
p^3_{\rm Anti.} &= {2 \pi (n + 1/2) \over L}\cr}\eqno(3)$$
The four dimensional momentum integration will now become a three
dimensional momentum integration and a discrete sum.  But for simplicity,
we will continue to use the standard notation of a four dimensional
integration with the understanding that
$$\int {d^4 p \over (2 \pi)^4} \equiv \int {d^3 p \over (2 \pi )^3} \cdot
 {1 \over  L}\ \sum_n \eqno(4)$$

With these basics, let us note that the generating functional for the free
Wess-Zumino theory (when $g=0$) is simply the product
$$Z = Z_B Z_F \eqno(5)$$
where the bosonic and the fermionic generating functionals
 are given by
$$\eqalign{Z_B &= \big[ \det
\big( \partial^2 +m^2 \big) \big]^{-1}\cr
Z_F &= \det \big( \partial^2 +m^2 \big) \cr}\eqno(6)$$
When the bosons and the fermions satisfy the same boundary conditions, each
of these determinants is evaluated on the same space of functions so that
the generating functional in (5) becomes unity.  This shows that the vacuum
energy for the free theory vanishes on this space
 only when both the bosonic and the
fermionic fields satisfy the same boundary condition.  When interactions
are present, the lowest order vacuum diagrams will be at two loop and have
the form
$$\eqalign{\qquad\qquad\qquad\qquad &= 4 im^2g^2 \int {d^4 k \over (2 \pi
)^4}\
{d^4p \over (2 \pi )^4} \
{1 \over (k^2 -m^2)( p^2-m^2)
((k-p)^2 - m^2)}\cr
\noalign{\vskip 18pt}%
\qquad\qquad\qquad\qquad &= -8 ig^2 \int {d^4 k \over (2 \pi
)^4}\
{d^4p \over (2 \pi )^4} \
{k \cdot p \over (k^2 -m^2) (p^2-m^2)
((k-p)^2 - m^2)}\cr
\noalign{\vskip 18pt}%
\qquad\qquad\qquad\qquad &= 4 ig^2 \int {d^4 k \over (2 \pi
)^4}\
{d^4p \over (2 \pi )^4} \
{1 \over (k^2 -m^2)( p^2-m^2)}\cr}\eqno(7)$$
It is a simple matter to check and see that these diagrams add up to zero
algebraically only when both the bosonic and the fermionic fields satisfy the
same boundary condition.  The analysis can be carried out completely
analogous to the conventional analysis
 order by order and it can be shown that the theory
does not develop a vacuum energy when both bosonic and fermionic fields
satisfy the same boundary condition.  This observation may be quite helpful
in the study of string compactification which preserves supersymmetry.

In a similar manner, one can show that the theory does not generate any
tadpoles.  Here we just give one example.
$$\eqalign{\qquad\qquad\qquad &= 3 mg \int {d^4 p \over (2 \pi)^4} \
{1 \over p^2 -m^2}\cr
\noalign{\vskip 14pt}%
\qquad\qquad\qquad &= mg \int {d^4 p \over (2 \pi )^4}\ {1 \over p^2 -m^2}\cr
\noalign{\vskip 14pt}%
\qquad\qquad\qquad &= -4mg \int {d^4 p \over (2 \pi )^4}
\ {1 \over p^2 -m^2}\cr}\eqno(8)$$
Once again, we see that for the same boundary condition, these terms add up
to zero.

Next, let us evaluate the two point functions at one loop.  Let us
concentrate on the $A$-field and note that when $A,\ B$ and
$\psi$ satisfy the same boundary condition, we can write

\ \ \ \

\vskip 1truein

$$\eqalignno{= - i \pi_A (k) &= 4g^2 \int {d^4 p \over (2 \pi)^4}\
{2 k \cdot (k+p) + m^2 \over
(p^2 -m^2)((k +p)^2 -m^2 )}\cr
&= 4 g^2 (k^2 + m^2 )\int {d^4 p \over (2 \pi)^4}\
{1 \over (p^2 -m^2)((k+p)^2 -m^2)}&(9)\cr}$$
where $k^\mu$ represents the external momentum.  We write a four vector
$p^\mu = (\hat {\bf p} , p^3)$ and note that $p^3 = {2 \pi n \over L}$ for
periodic boundary condition whereas $p^3 = {2 \pi (n + {1 \over 2}) \over L
}$ for antiperiodic boundary condition (see Eq. (3)).  Let us first
evaluate the self-energy in (9) for periodic boundary condition.  Rotating
to Euclidean space, we obtain
$$\eqalign{-i &\pi^{\rm Per.}_A (k_E) = -4ig^2 \big( k^2_E - m^2 \big)
\int {d^4 p_E \over (2 \pi)^4}\ {1 \over (p^2_E +m^2)((k_E +p_E)^2
+m^2)}\cr
&= -4 ig^2 \big( k^2_E -m^2 \big) \int
{d^3 \hat p_E \over (2 \pi)^3} \ {1 \over  L} \sum_n
{1 \over (\hat {\bf p}^2_E +({2 \pi n \over L})^2 + m^2 )
((\hat {\bf k}_E + \hat {\bf p}_E )^2 + (k^3 +
{2 \pi n \over L})^2 +
m^2)}\cr}\eqno(10)$$
The sum over $n$ can be performed using the formula
$$\sum_n f(n) = - \sum {\rm Res}\ f(z) \pi \cot  \pi z
\qquad {\rm  at\  the\  poles\  of}\
f(z)\eqno(11)$$
for any function $f(z) \rightarrow 0$ as $|z| \rightarrow \infty$.  We also
note that
$$\eqalign{\coth {z \over 2} &= {e^{z/2} + e^{-z/2} \over
e^{z/2} - e^{-z/2}}\cr
&= (1 + 2n_B (z))\cr}\eqno(12)$$
where
$$n_B (z) = {1 \over e^z -1} \eqno(13)$$
represents the bosonic distribution function.  Using Eqs. (11) and (12),
the sum in Eq. (10) can be performed and the result has the form
$$-i \pi^{\rm Per.}_A (k_E) = -i \pi^{(0)}_A (k_E) - i
\pi^{\rm Per.}_{A,L} (k_E) \eqno(14)$$
where $\pi^{(0)}_A (k_E)$ represents the $L$-independent self-energy of the
supersymmetric theory in the conventional Euclidean space (namely, $L
\rightarrow \infty$ limit) and
$$\eqalignno{-i \pi^{\rm Per.}_{A,L}(k_E) &= -4ig^2 \big( k^2_E -m^2 \big)
\int {d^3 \hat p_E \over (2 \pi )^3}
\ {n_B (L \omega_{\hat p}) \over \omega_{\hat p}} \times\cr
&\qquad \bigg[ {1 \over k^2_E +2 (\hat {\bf k}_E \cdot
\hat {\bf p}_E +i k^3 \omega_{\hat p})}
+ {1 \over k^2_E +2 (\hat {\bf k}_E \cdot \hat
{\bf p}_E - ik^3
\omega_p)}\bigg]\cr
\noalign{\vskip 5pt}%
&= -{i g^2 \over 2 \pi^2}\ {(k^2_E -m^2 )\over \hat k_E}
\int^\infty_0 d \hat p_E {\hat p_E n_B (L \omega_{\hat p}) \over
\omega_{\hat p}} \ \ln R &(15)\cr}$$
where we have defined
$$\eqalign{\omega_{\hat p} &= \big( \hat p^2_E + m^2 \big)^{1/2}\cr
R &= {\big( k^2_E + 2 \big( \hat k_E \hat p_E + ik^3
\omega_{\hat p} \big) \big) \big( k^2_E + 2 \big(
\hat k_E \hat p_E -ik^3 \omega_{\hat p} \big) \big) \over
\big( k^2_E -2 \big( \hat k_E \hat p_E + i k^3 \omega_{\hat p}
\big) \big) \big( k^2_E -2 \big( \hat k_E \hat p_E -
i k^3 \omega_{\hat p} \big)\big)}\cr}
\eqno(16)$$
and $\hat k_E$ and $\hat p_E$ in Eqs. (15) and (16) stand for the lengths
 of the Euclidian three vectors $\hat {\bf k}_E$ and $\hat {\bf p}_E$.

We can, similarly, evaluate the self-energy for the $A$-field with
antiperiodic boundary condition for all the fields.  The relevant formula
to use for the evaluation of the discrete sum, in this case, is
$$\sum f (2n +1) = \sum \ {\rm Res}\ f (z) \pi \tan \pi z \qquad
{\rm at\ the\  poles\ of}\ f(z) \eqno(17)$$
and the result is
$$-i \pi^{\rm Anti.}_A (k_E) = - i \pi^{(0)}_A (k_E) -
i \pi^{\rm Anti.}_{A,L} (k_E)\eqno(18)$$
with
$$-i \pi^{Anti.}_{A,L} (k_E) = {ig^2 \over 2 \pi^2}\
{(k^2_E -m^2) \over \hat k_E} \int^\infty_0 d \hat p_E
\ {\hat p_E n_F (L \omega_{\hat p}) \over
\omega_{\hat p}} \ \ln R \eqno(19)$$
where
$$n_F (z) = {1 \over e^z + 1} \eqno(20)$$
defines the Fermi distribution function.

The self-energies for the $B$-field and the $\psi$-field can be, similarly,
evaluated.  We note here the results without going into details.

\ \ \ \

\vskip 1truein

$$= - i \pi_B (k) = - i \pi^{(0)}_B (k) - i
\pi_{B,L} (k) \eqno(21)$$
where
$$\eqalignno{-i \pi^{\rm Per.}_{B,L} (k_E) &= -{ig^2 \over 2 \pi^2}\
{(k^2_E -m^2) \over \hat k_E} \int^\infty_0 d \hat p_E
\ {\hat p_E n_B (L \omega_{\hat p}) \over \omega_{\hat p}}\ \ln R&(22)\cr
-i \pi^{\rm Anti.}_{B,L} (k_E) &= {ig^2 \over 2 \pi^2}\
{(k^2_E -m^2) \over \hat k_E} \int^\infty_0 d \hat p_E
\ {\hat p_E n_F (L \omega_{\hat p}) \over \omega_{\hat p}}\ \ln R&(23)\cr}$$
Similarly, for the $\psi$-field we have

\ \ \ \

\vskip 1 truein

$$= - i \pi_\psi (k) = - i \pi^{(0)}_\psi (k) -i \pi_{\psi , L} (k)
\eqno(24)$$
with
$$\eqalignno{-i \pi^{\rm Per.}_{\psi , L} (k_E) &= -{i g^2 \over
2 \pi^2}\ {\rlap \slash{k}_E \over \hat k_E} \int^\infty_0 d
\hat p_E \ {\hat p_E n_B (L \omega_{\hat p})\over
\omega_{\hat p}} \ \ln R&(25)\cr
-i \pi^{\rm Anti.}_{\psi , L} (k_E) &= {i g^2 \over
2 \pi^2}\ {\rlap \slash{k}_E \over \hat k_E} \int^\infty_0 d
\hat p_E \ {\hat p_E n_F (L \omega_{\hat p})\over
\omega_{\hat p}} \ \ln R&(26)\cr}$$

We note from Eqs. (15), (19), (22), (23), (25) and (26) that when
all the fields satisfy the same boundary condition, we can write
$$\eqalign{-i \pi_{A,L} (k_E) &= \big( k^2_E -m^2 \big) \pi_L
(k_E)\cr
-i \pi_{B,L} (k_E) &= \big( k^2_E -m^2 \big) \pi_L
(k_E)\cr
-i \pi_\psi (k_E) &= \rlap \slash{k}_E  \pi_L (k_E)\cr}\eqno(27)$$
where
$$\eqalign{\pi^{\rm Per.}_L (k_E) &= -{ig^2 \over 2 \pi^2 \hat k_E}
\int^\infty_0 d \hat p_E \ {\hat p_E n_B (L \omega_{\hat p})
\over \omega_{\hat p}} \ \ln R\cr
\pi^{\rm Anti.}_L (k_E) &= {ig^2 \over 2 \pi^2 \hat k_E}
\int^\infty_0 d \hat p_E \ {\hat p_E n_F (L \omega_{\hat p})
\over \omega_{\hat p}} \ \ln R\cr}\eqno(28)$$
Once again we see that as long as all the fields satisfy the same boundary
condition, the form of the one loop self-energies is consistent with the
nonrenormalization theorem (see Eq. (2)).  However, we note that, in this
case, the self-energies become nonanalytic.  It is easy to see that if we
set $k^3 = 0$ and then take $\hat k_E \rightarrow 0$, we obtain
$$\eqalign{\lim_{\hat k_E \rightarrow 0 \ k^3 \rightarrow 0}
\pi^{\rm Per.}_L (k_E)
&= -{ig^2 \over  \pi^2} \int^\infty_0 d \hat p_E
{n_B (L \omega_{\hat p})\over
\omega_{\hat p}}\cr
\lim_{\hat k_E \rightarrow 0\ k^3 \rightarrow
 0} \pi^{\rm Anti.}_L (k_E)
&= {ig^2 \over  \pi^2} \int^\infty_0 d \hat p_E
{n_F (L \omega_{\hat p})\over
\omega_{\hat p}}\cr}\eqno(29)$$
On the other hand, if we set $\hat k_E =0$
 and then take the limit $k^3 \rightarrow 0$, this yields
$$\eqalign{\lim_{ k^3 \rightarrow 0\ \hat k_E
\rightarrow 0} \pi^{\rm Per.}_L (k_E)
 &= -{ig^2 \over  \pi^2} \int^\infty_0 d \hat p_E
{\hat p_E^2 n_B (L \omega_{\hat p})\over
\omega_{\hat p}^3}\cr
\lim_{ k^3 \rightarrow 0\ \hat k_E \rightarrow 0 } \pi^{\rm Anti.}_L (k_E)
&= {ig^2 \over  \pi^2} \int^\infty_0 d \hat p_E
{\hat p_E^2 n_F (L \omega_{\hat p})\over
\omega_{\hat p}^3}\cr}\eqno(30)$$
This is exactly the same nonanalyticity seen in the context of finite
temperature [1,3] and supersymmetry has not improved this behavior.  In fact,
supersymmetry only seems to imply that the nonanalyticity in the bosonic
and the fermionic two point functions are the same. Even though we have
studied only the Wess-Zumino theory, a little analysis would indicate that
the nonanalyticity in the two point functions would be present in any
supersymmetric theory on ${\bf R}^3 \times S^1$ simply because, the fermion
self-energy graphs are not of cancelling nature.

In closing, we note that periodic boundary condition for fermions is known
to lead to problematic causal behavior for
 propagators [9].  This can be easily
seen from the fact that the quantum corrections to the two point functions
have opposite sign corresponding to periodic or antiperiodic boundary
conditions.  We have not particularly worried as to which boundary condition
would be physical in such theories.  Rather, we  have shown that either of
the boundary conditions seem to be consistent with supersymmetry even
though they may lead to nonanalytic amplitudes.  It may very well be that
for such theories the only physical boundary condition is when all fields
are antiperiodic in the compact direction. This needs further study, in
particular, if we are to take compactification seriously.

This work was supported in part by U.S. Department of Energy Grant No.
DE-FG-02-91ER40685.
    M.H. would like to thank the
Funda\c c\~ao de Amparo a Pesquisa do Estado de S\~ao
 Paulo for the financial
support.

\vfill\eject

\noindent {\bf \underbar{References}}

\item{1.} H.A. Weldon, Phys. Rev. {\bf D26} (1982) 1394; ibid {\bf D28}
 (1983) 2007.

\item{2.} For a review of the problem and references see P.S. Gribosky and
B.R. Holstein, Z. Phys. {\bf C47} (1990) 205;
 P.F. Bedaque and A. Das, Phys. Rev. {\bf D45} (1992) 2906.

\item{3.} H.A. Weldon, Phys. Rev. {\bf D47} (1993) 594; P.F. Bedaque and A.
Das, Phys. Rev. {\bf D47} (1993) 601.

\item{4.} P. Arnold, S. Vokos, P. F. Bedaque and A. Das, Phys. Rev. {\bf
D47} (1993) 498.

\item{5.} A. Das and M. Kaku, Phys. Rev. {\bf D18} (1978) 4540.

\item{6.} A. Das, Physica A158 (1989) 1.

\item{7.} L.H. Ford and T. Yoshimura, Phys. Lett. {\bf A70}
 (1979) 89.

\item{8.} B.S. Kay, Phys. Rev. {\bf D20} (1979) 3052.

\item{9.} L.H. Ford, Phys. Rev. {\bf D20} (1980) 933.

\item{10.} L.H. Ford and N.F. Svaiter, \lq\lq One-Loop Renormalization of
Self-interacting Scalar Fields in Non simply Connected Space-times",
TUTP-94-18, hep-th/9411109.

\item{11.} G. Denardo and E. Spallucci, Nucl. Phys. {\bf B169} (1980) 514.

\item{12.} D. J. Toms, Phys. Rev. {\bf D21} (1980) 928.

\item{13.} R. Kubo, J. Phys. Soc. Japan 12 (1957) 570; P. Martin and J.
Schwinger, Phys. Rev. 115 (1959) 1342.

\item{14.} See, for example, \lq\lq Superspace or One Thousand and One
Lessons in Supersymmetry", S.J. Gates, Jr., M. Grisaru, M. Rocek and W.
Siegel, Benjamin/Cummings, Reading 1983.

\end